\documentclass[english,prd,superscriptaddress,nofootinbib,preprintnumbers,showpacs,preprint]{revtex4}
\usepackage[latin1]{inputenc}
\usepackage{graphicx}
\usepackage{amsmath,amsthm,amssymb}
\usepackage{bm}

\usepackage{amsfonts}
\usepackage{dcolumn}

\usepackage{color}

\newcommand{\hatR}{\widehat{R}}
\newcommand{\hatg}{\widehat{g}}
\newcommand{\hatnabla}{\widehat{\nabla}}
\newcommand{\hatV}{\widehat{V}}
\newcommand{\hatphi}{\widehat{\phi}}
\newcommand{\hatepsilon}{\widehat{\epsilon}}
\newcommand{\hateta}{\widehat{\eta}}
\newcommand{\hatxi}{\widehat{\xi}}

\newcommand{\beqa}{\begin{eqnarray}}
\newcommand{\eeqa}{\end{eqnarray}}

\newcommand{\siml}{\lesssim}

\usepackage{babel}
\begin{document}

\title{Consistency Relations for Large Field Inflation: Non-minimal Coupling
}

\author{Takeshi Chiba}
\affiliation{Department of Physics, College of Humanities and Sciences, 
Nihon University, Tokyo 156-8550, Japan}
\author{Kazunori Kohri}
\affiliation{Institute of Particles and Nuclear Studies, KEK, and Sokendai, Tsukuba 305-0801, Japan}

\begin{abstract}
We derive the consistency relations for  a chaotic inflation model 
with a non-minimal coupling to gravity.  For a quadratic  potential 
 in the limit of a small non-minimal coupling parameter $\xi$ and for a quartic 
potential without assuming small $\xi$, we give 
the consistency relations among the spectral index $n_s$, 
the tensor-to-scalar ratio $r$ and the running of the spectral index 
$\alpha$. 
We find that unlike $r$, $\alpha$ is less sensitive to $\xi$. 
If $r<0.1$,  then $\xi$ is constrained to  $\xi<0$  and 
$\alpha$ is predicted to be  $\alpha\simeq -8\times 10^{-4}$ 
for a quartic potential. For a general monomial potential, $\alpha$ is constrained in the range  $-2.7\times 10^{-3}<\alpha< -6\times 10^{-4}$ as long as $|\xi|\leq 10^{-3}$ 
if $r<0.1$. 

\end{abstract}

\date{\today}


\maketitle

\section{Introduction}

In our previous paper, motivated by the possibility of a large  tensor-to-scalar ratio $r$ \cite{bicep2}, we provided the consistency relations among 
the spectral index $n_s$, the tensor-to-scalar ratio $r$, and the running of the spectral index 
$\alpha$ for several large field inflation models (chaotic with monomial potential, natural, 
symmetry breaking) \cite{ck}. The basic idea is to construct  
one relation out of two model parameters using three observables ($n_s, r$, and $\alpha$). 
We find that $\alpha$ can be a discriminating probe of large field inflation models. 

In this paper, we investigate the stability of the consistency relation for  
chaotic inflation with a monomial potential  that  we have recently found. 
To do this, we consider a non-minimal coupling as a "perturber" of the model. 
Then, the number of model parameters becomes three and we need a fourth observable (for example, the "running" of $\alpha$),  but this would introduce complication and 
the comparison with the "unperturbed" relation would be difficult. So, in this paper 
we fix one of the model parameters and examine how introducing  the non-minimal 
coupling affects the relation.

\section{Consistency Relations for Chaotic Inflation with a Non-minimal Coupling}
\label{sec2}

\subsection{From Jordan to Einstein}

We consider a single field inflation model 
with a non-minimal coupling to gravity. The action is given by
\beqa
S=\int\sqrt{-g}\left(\frac{1}{16\pi G_*} \Omega(\phi)R-\frac12(\nabla\phi)^2 -V(\phi)\right)\,,
\eeqa
where $g_{\mu\nu}$ is the Jordan frame metric and $G_*$ is the bare gravitational constant and we shall set 
$8\pi G_*=1$ henceforth. 
As $\Omega(\phi)$ and $V(\phi)$, we take
\beqa 
\Omega(\phi)=1-\xi\phi^2,\qquad V(\phi)=\frac{\lambda}{n} \phi^n\,,
\label{omega-v}
\eeqa
where $\xi$ is a non-minimal coupling parameter. In our convention, $\xi=1/6$ corresponds to 
the conformal coupling. 

As is well known, by introducing the new metric 
called Einstein frame metric $\hatg_{\mu\nu}=\Omega g_{\mu\nu}$, the action
 can be rewritten as that 
of Einstein gravity with a scalar field \cite{fm}:
\beqa
S=\int\sqrt{-\hatg}\left( \frac12\hatR-\frac{1}{2\Omega}
\left(1+\frac{3\Omega_{,\phi}^2}{2\Omega}\right)(\hatnabla\phi)^2-\frac{V}{\Omega^2} \right)\,,
\eeqa  
where the hatted variables are defined by $\hatg_{\mu\nu}$ and  $\Omega_{,\phi}=d\Omega/d\phi$. Hence in terms of the canonically normalized scalar field $\hat\phi$ defined by
\beqa
d\hat\phi^2=\frac{1}{\Omega(\phi)}\left(1+\frac{3\Omega(\phi)_{,\phi}^2}{2\Omega(\phi)}\right)d\phi^2\equiv\frac{f(\phi)}{\Omega(\phi)}\,,
\label{hatphi-phi}
\eeqa
the system is reduced to the Einstein gravity plus 
a minimally coupled scalar field with the effective 
potential $\hatV$ defined by
\beqa
\hatV=\frac{V}{\Omega^2}\,.
\eeqa
For $\Omega$ and $V$ in Eq. (\ref{omega-v}), $\hatV$ with $n=4$  becomes flat for large $|\xi\phi^2|$ with $\xi<0$, 
which is the essence of the Higgs \cite{higgs} (or Starobinsky \cite{alex}) inflation.

\subsection{$r,n_s$, and $\alpha$}

Hence, in order to compute the spectral index $n_s$, 
the tensor-to-scalar ratio $r$ and the running of the spectral index 
$\alpha$, we only need to calculate slow-roll parameters in terms of $\hatphi$ and $\hatV$:
\beqa
\hatepsilon\equiv \frac12\left(\frac{\hatV_{,\hatphi}}{\hatV}\right)^2,  \quad 
\hateta\equiv\frac{\hatV_{,\hatphi\hatphi}}{\hatV},  
\quad 
\hatxi\equiv \frac{\hatV_{,\hatphi}\hatV_{,\hatphi\hatphi\hatphi}}{\hatV^2} \,.
\eeqa
Then $r, n_s$, and $\alpha$ are given by 
\beqa
r=16\hatepsilon,  \quad n_s-1=-6\hatepsilon+2\hateta, \quad \alpha=16\hatepsilon\hateta-24\hatepsilon^2-2\hatxi
\eeqa
In fact, for a single scalar field, the observables are 
independent of the conformal transformation \cite{ms,cy}.

For example, in the limit of small $\xi$, the slow-roll parameters become
\beqa
\hatepsilon&=&\frac{n^2}{2 \phi ^2}-\frac{1}{2} ((n-8) n) \xi ,\\
\hateta&=&\frac{(n-1) n}{\phi ^2}+(4-(n-8) n) \xi  ,\\
\hatxi&=&\frac{(n-2) (n-1) n^2}{\phi ^4}+\frac{n (n ((19-2 n) n-14)+8) \xi
   }{\phi ^2}+O(\xi^2\phi^{-2})\,,
\eeqa
and $r,n_s$, and $\alpha$ are given by
\beqa
r&=&\frac{8 n^2}{\phi ^2}-8 ((n-8) n) \xi,
\label{smallxi1} \\
n_s-1&=&-\frac{n (n+2)}{\phi ^2}+((n-8) n+8) \xi,
\label{smallxi2} \\
\alpha&=&-\frac{2 \left(n^2 (n+2)\right)}{\phi ^4}+
\frac{2 (n-4) n (n+2) \xi   }{\phi ^2}+16n \xi^2+O(\xi^2\phi^{-2})
\label{smallxi3}
\eeqa
On the other hand, for $n=4$ with large $|\xi|\phi^2$, we have 
\beqa
\hatepsilon&=&-\frac{8}{\left(1-6 \xi \right) \xi\phi
   ^4},\\
\hateta&=&-\frac{8}{(1-6 \xi) \phi ^2}+\frac{4 (12 \xi -5)}{(1-6 \xi )^2 \xi 
   \phi ^4},\\
\hatxi&=&\frac{64}{(1-6 \xi )^2 \phi ^4}+\frac{64 (7-18 \xi )}{  (1-6 \xi
   )^3 \xi\phi ^6} ,
\eeqa
and 
\beqa
r&=&-\frac{128}{(1-6\xi)\xi\phi^4},
\label{largexi1}\\
n_s-1&=&-\frac{16}{(1-6\xi)\phi^2}+\frac{8(1-24\xi)}{(1-6\xi)^2\xi\phi^4},
\label{largexi2}\\
\alpha&=&-\frac{128}{(1-6\xi)^2\phi^4}+\frac{128(1-30\xi)}{(1-6\xi)^3\xi\phi^6}\,.
\label{largexi3}
\eeqa

\subsection{e-folding number}

Finally, we  provide the relation for the e-folding number until the end of inflation  
$N$. Since the scale factor and the proper time  in the Jordan frame $a$ and $t$ are 
related to those in   the Einstein frame $a=\Omega^{1/2}\widehat{a}$ and $dt=\Omega^{1/2}d\widehat{t}$,   
the Hubble parameter  in the Jordan frame $H$  
is related to that in the Einstein frame $\widehat{H}$  
by the relation \cite{catena, cy2}
\beqa
H=\frac{da/dt}{a}=\frac{1}{\Omega^{1/2}}\left(\widehat{H}+\frac{d\Omega/d\widehat{t}}{2\Omega}\right),
\eeqa
and the e-folding number $N$ is given by
\beqa
N=\int^{t_{end}} Hdt=\int^{\widehat{t}_{end}}\widehat{H}d\widehat{t}+\frac12\int_{\hatphi_{end}}\frac{\Omega_{,\hatphi}}{\Omega}d\hatphi \,.
\eeqa
Under the slow-roll approximation, $|\ddot\phi| \ll H|\dot\phi|$ and 
$|\dot\Omega|\ll H\Omega$,    using the slow-roll equations of motion  \cite{cy2}
\beqa
3H\dot\phi\simeq -\frac{\Omega^2}{f}\left(\frac{V}{\Omega^2}\right)_{,\phi}\,, 
  \quad 3H^2\Omega\simeq V \,,
\eeqa
$N$ becomes
\beqa
N \simeq \int_{\phi_{end}} 
\frac{f V}{\Omega^3\left(V/\Omega^2\right)_{,\phi}}d\phi=\int_{\hatphi_{end}} 
\frac{\hatV}{\hatV_{,\hatphi}}d\hatphi \,, 
\eeqa
where $f(\phi)$ is defined by Eq. (\ref{hatphi-phi}). Note that $\widehat{H}+d\Omega/d\widehat{t}/(2\Omega)$ is the Hubble parameter in 
the Einstein frame that measures the distance \cite{catena, cy2}.  
The e-folding number 
is frame-invariant and  can be calculated in either frame. 
For example, for $|\xi|\ll1$, $N$ is given by $N\simeq \phi^2/(2n)$, and for 
$n=4$ and $|\xi|\phi^2\gg 1$, $N\simeq (1-6\xi)\phi^2/8$.

\section{Consistency Relations}

\subsection{$|\xi|\ll 1$}

Given the series expansion Eqs. (\ref{smallxi1})-(\ref{smallxi3}) for $|\xi|\ll 1$, 
we may rewrite $\alpha$ in terms of $r$ and $n_s$ for fixed $n$. 
We consider the quadratic ($n=2$) case and the quartic ($n=4$) case, respectively.

\subsubsection{  $n=2$} 

For $n=2$, Eqs. (\ref{smallxi1})-(\ref{smallxi3}) become
\beqa
r&=&\frac{32}{\phi^2}+96\xi,\\
n_s-1&=&-\frac{8}{\phi^2}-4\xi,\\
\alpha&=&-\frac{32}{\phi^4}-\frac{32\xi}{\phi^2}
+32\xi^2\,,
\eeqa
and we find a consistency relation for $n=2$ with $|\xi|\ll 1$:
\beqa
&&\alpha=\frac{1}{160}\left(r+4(n_s-1)\right)^2-\frac12 (n_s-1)^2,
\label{alpha:n2}\\
&&r<24(1-n_s)\,,
\eeqa
where the second inequality follows from the positivity of $\phi^2$.  
Interestingly, the minimum of $\alpha$ is achieved at $r=4(1-n_s)$ with 
$\alpha=-(1/2)(n_s-1)^2=-(1/32)r^2$, which coincides with 
the relation for the minimally coupled  ($\xi=0$) scalar field \cite{ck}.  
Since $\xi=0$ is the minimum (extremum) of $\alpha$,  $\alpha$ is insensitive to $\xi$. 
$r$ is written as $r= 4(1-n_s) +80\xi$.  So, the expansion is valid for $|\xi|\siml 10^{-3}$. 
In fact, as shown in Fig. \ref{fig10},  the expansion is accurate within  $O(10) \%$ 
 for $|\xi|\siml 10^{-3}$.  We note that the current observational constraint on $\xi$ is $-5.1\times 10^{-3}<\xi\leq 0$ \cite{tsujikawa}.

\subsubsection{ $n=4$} 

\begin{figure}
\includegraphics[height=3.3in,width=5.0in]{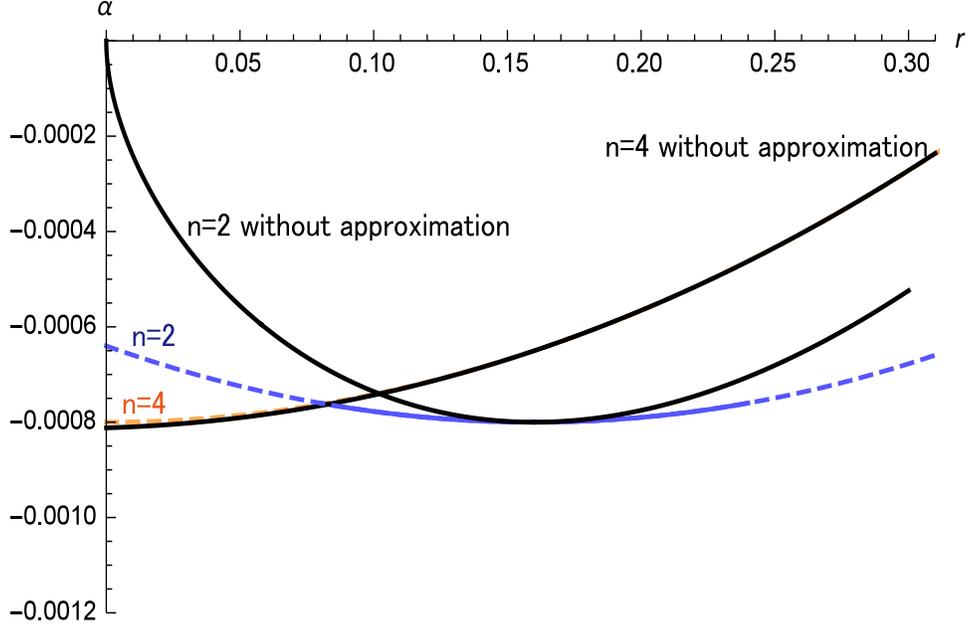}
\caption{\label{fig10}
The relation between $r$ and $\alpha$ for $n_s=0.96$.  The dashed blue curve is 
Eq.  (\ref{alpha:n2}) derived assuming $|\xi|\ll 1$, while the upper left black one is the curve 
 derived without assuming   $|\xi|\ll 1$. 
The dashed orange curve is 
Eq. (\ref{alpha:n4}) derived assuming $|\xi|\ll 1$,  
which almost overlaps with  the black one derived
without assuming $|\xi|\ll 1$.
 }
\end{figure}

For $n=4$, Eqs. (\ref{smallxi1})-(\ref{smallxi3}) become 
\beqa
r&=&\frac{128}{\phi^2}+128\xi,\\
n_s-1&=&-\frac{24}{\phi^2}-8\xi,\\
\alpha&=&-\frac{192}{\phi^4}+64\xi^2\,,
\eeqa
We find a consistency relation for $n=4$ with $|\xi|\ll 1$:
\beqa
&&\alpha=\frac{3}{512}r^2-\frac12 (n_s-1)^2,
\label{alpha:n4}\\
&&r<16(1-n_s)\,,
\eeqa
where again the second inequality follows 
from the positivity of $\phi^2$. $\xi=0$ corresponds to 
$r=(16/3)(1-n_s)$ \cite{ck}.  
Although the expansion is apparently valid for $|\xi|\siml10^{-3}$, 
we find that Eq. (\ref{alpha:n4}) fits extremely well with the curve without assuming  small $|\xi|$ (see Fig. \ref{fig10}). 
The current observational  constraint on $\xi$ is $\xi<-1.9\times 10^{-3}$ \cite{planck22,tsujikawa}.

\subsubsection{$n (\neq 4)$}

For general $n(\neq 4)$, from Eqs. (\ref{smallxi1})-(\ref{smallxi3}),  
we obtain
\beqa
&&\alpha=
\frac{(n+2) \left(3 n^3-32 n^2+88 n-64\right)}{64 (n-12)^2 n^2}r^2
+\frac{(n+2) \left(5 n^2-48 n+80\right)} {8 (n-12)^2 n}r (n_s-1)
\nonumber\\
&&\qquad +\frac{2(n+2) (n-8) }{ (n-12)^2 } (n_s-1)^2,\\
&&(12-n)\left( \left(8(n-1)-n^2\right)r  -8n(8-n)(1-n_s)\right)<0
\,.
\eeqa

\subsection{$n=4$ with $|\xi|\gg 1$}

For $n=4$ with $|\xi|\gg 1$, Eqs. (\ref{largexi1})-(\ref{largexi3}) 
become \footnote{$\alpha$ in the Higgs inflation was calculated in \cite{hkop}.}
\beqa
r&=&\frac{64}{3\xi^2\phi^4},\label{largexir}\\
n_s-1&=&\frac{8}{3\xi\phi^2},\label{largexin}\\
\alpha &=&-\frac{32}{9\xi^2\phi^4}\,.
\label{largexia}
\eeqa 
Hence we obtain
\beqa
\alpha=-\frac12(n_s-1)^2=-\frac16 r \,,
\label{alpha:largexi:n4}
\eeqa 
which may be called "Starobinsky attractor" according to \cite{klr}. \footnote{This large 
$|\xi|$ behavior can be generalized by replacing $\xi\phi^2$ with $\xi g(\phi)$ 
so that  $\Omega(\phi)=1-\xi g(\phi)$ and $V(\phi)=\lambda g(\phi)^2$. \cite{klr}}

\begin{figure}
\includegraphics[height=3.3in,width=5.0in]{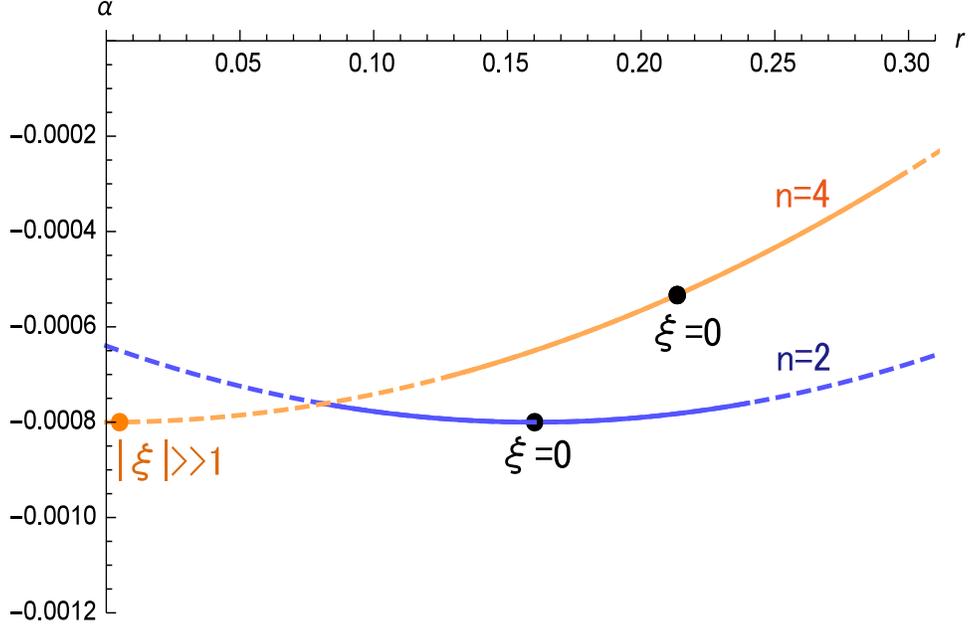}
\caption{\label{fig1}
Consistency relations for the nonminimally coupled chaotic inflation 
model with $n=2$ (dashed blue) and $n=4$ (dashed orange) for $n_s=0.96$ 
in $(r,\alpha)$ plane. Black points are  for $\xi=0$. 
Solid curves are for $|\xi|<10^{-3}$. 
Orange point is for $|\xi|\gg 1$ (Eq. (\ref{alpha:largexi:n4})). Note that the dashed 
blue curve is inaccurate. 
 }
\end{figure}

\begin{figure}
\includegraphics[height=3.3in,width=5.0in]{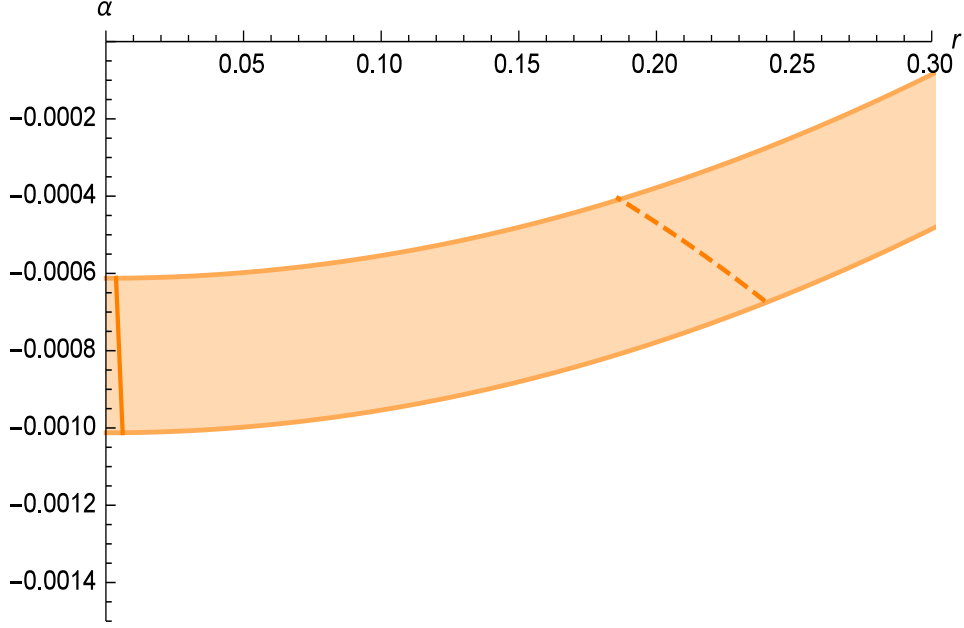}
\caption{\label{fig11}
Consistency relations for the nonminimally coupled chaotic inflation 
model with a quartic potential  for $0.955<n_s<0.965$ in $(r,\alpha)$ plane. 
Dashed curves are for $\xi=0$. Solid orange curve is for $|\xi|\gg 1$ 
(Eq. (\ref{alpha:largexi:n4})). 
 }
\end{figure}

In Fig. \ref{fig1}, we show the relations 
Eq. (\ref{alpha:n2}), Eq. (\ref{alpha:n4}), and Eq. (\ref{alpha:largexi:n4}) in the $(r,\alpha)$ 
plane for $n_s=0.96$. 
Black points are  for $\xi=0$, the left (right) of which corresponds to  $\xi<0 (\xi>0)$. 
Solid curves are for $|\xi|<10^{-3}$. 
Orange point is for $|\xi|\gg 1$ (Eq. (\ref{alpha:largexi:n4})). 
Fig. \ref{fig11} shows the regions scanned by 
the relations Eq. (\ref{alpha:n4}) and Eq. (\ref{alpha:largexi:n4}) for  $0.955<n_s<0.965$. 
We find that $\alpha$ is insensitive to $\xi$. For $\xi<0$,  $\alpha$ is constrained in 
the narrow range:  $-8 \times 10^{-4}<\alpha<-6 \times 10^{-4}$ for $n_s=0.96$, and  
$- 10^{-3}<\alpha<-4 \times 10^{-4}$ for $0.955<n_s<0.965$. \footnote{We note that 
the e-folding number $N$ for higher $n_s$ can exceed the standard upper limit $N<60$ 
\cite{dh}, which may require non-standard thermal history of the universe \cite{ll}.}


\subsection{General  $n$ for fixed $\xi$.}

Lastly, we provide a consistency relation for general $n$ for 
fixed $\xi$ with $|\xi| \ll 1$. 
{}From Eqs. (\ref{smallxi1})-(\ref{smallxi2}), $n$ and $\phi^2$ are written in terms of $r$, 
 $n_s$, and $\xi$ :
\beqa
n&=&\frac{192 \xi -r-8(n_s-1)-\sqrt{(r+8 (n_s-1-24 \xi ))^2-128 \xi  r}}{32 \xi },\\
\phi^2&=&\frac{r+8(n_s-1)-192 \xi +\sqrt{(r+8 (n_s-1-24 \xi ))^2-128 \xi  r}}
{2 \xi  (r+8   (n_s-1-8 \xi ))}.
\eeqa
Then, from Eq. (\ref{smallxi3}), $\alpha$ can be written as a function of $r$ and $n_s$,  
which is too complicated to show here. 

\begin{figure}
\includegraphics[height=3.3in,width=5.0in]{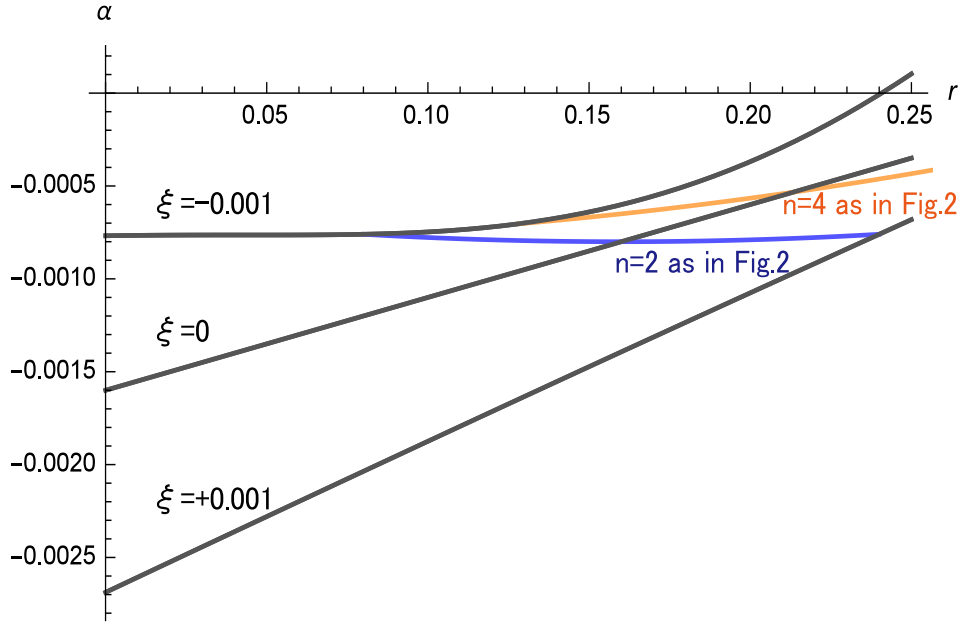}
\caption{\label{fig2}
Consistency relations for the nonminimally coupled chaotic inflation 
model with  $\xi=-10^{-3},0,10^{-3}$ from top to bottom for $n_s=0.96$.
Blue (orange) curve corresponds to $n=2(n=4)$ as in Fig. \ref{fig1}. 
}
\end{figure}

In Fig. \ref{fig2}, we plot $\alpha$ as a function of $r$ 
for $n_s=0.96$.   $\xi=-10^{-3},0,10^{-3}$ from top to bottom.  
The curves for $n=2$ and $n=4$ are also shown.  We find that 
the consistency relation for $\xi=0$ ($\alpha=-(1-n_s)^2 +\frac18 r(1-n_s)$ ) 
derived in \cite{ck} does not change so much 
as long as $|\xi|<10^{-3}$. 
For general $n$, varying $\xi$ changes $\alpha$ by $O(10^{-3})$. 
If  $r<0.1$, then   
$\alpha$ is constrained in the range of $-2.7\times 10^{-3}<\alpha< -8\times 10^{-4}$. 

%

\section{Summary} 
\label{secsum}

We have derived consistency relations for chaotic inflation with a nonminimal coupling  $\xi$.   
For a quadratic potential, 
we find that although the tensor-to-scalar ratio 
$r$ is sensitive to $\xi$, 
 the running of the spectral index $\alpha$ is rather 
insensitive to the change in  $\xi$ as long as $\xi $ is small.  
For a quartic potential, we find that $\alpha$ is insensitive to 
$\xi$  even for large $|\xi|$. 
We also find that the consistency relation for a general 
monomial potential does not change so much by changing $\xi$ 
as long as $|\xi|\leq 10^{-3}$. 

If  $r<0.1$, then $\xi<0$ and  
$\alpha\simeq -8 \times 10^{-4}$ are implied for a  quartic potential. 
Even for a general monomial potential, $r<0.1$ forces $\alpha$ to be in the range $-2.7\times 10^{-3}<\alpha< -8\times 10^{-4}$ for $n_s=0.96$ as long as $|\xi|<10^{-3}$ . 
Since  $\alpha$ is found to be insensitive to $\xi$,  
this $\alpha$ may be regarded as the prediction for the chaotic potential irrespective of the nonminimal coupling. 
Measurement of $\alpha$ with a precision of $O(10^{-3})$ 
by future observations of the 21 cm line \cite{kohri} will be crucially important 
in  pinning down the inflation model. 

Note added in proof: A recent joint analysis of BICEP2/Keck Array and Planck data 
yields an upper limit $r<0.12$ \cite{bicep2planck}.

\section*{ACKNOWLEDGEMENTS}

T.C. would like to thank Masahide Yamaguchi and Atsushi Naruko for useful comments. 
This work is supported by the Grant-in-Aid for Scientific Research
from JSPS (Nos.\,24540287 (TC), 23540327, 26105520 and 26247042 (KK)), and in part
by Nihon University (TC), and by the Center for the Promotion of
Integrated Science (CPIS) of Sokendai 1HB5804100 (KK).


\end{document}